\documentclass[aps,prb,twocolumn,groupedaddress,showpacs]{revtex4-1}  
\usepackage{graphics}
\usepackage{graphicx}
\usepackage{amsmath}
\usepackage{amssymb}
\usepackage{amsthm}
\usepackage{epsfig}
\usepackage{dsfont}
\usepackage{xcolor} %\textcolor{red}{}
\begin{document}
%%%%%%%%%%%%%%%%%%%%%%%%%%%%%%%%%%%%%%%%%%%%%%%%%%%%%%%%%%%%%%%%
\title{A Novel Treatment of the Josephson Effect}
%%%%%%%%%%%%%%%%%%%%%%%%%%%%%%%%%%%%%%%%%%%%%%%%%%%%%%%%%%%%%%%%
\author{Jacob Szeftel$^1$}
\email[corresponding author :\quad]{jszeftel@lpqm.ens-cachan.fr}
\author{Nicolas Sandeau$^2$}
\author{Michel Abou Ghantous$^3$}
\affiliation{$^1$ENS Paris-Saclay/LuMIn, 4 avenue des Sciences, 91190 Gif-sur-Yvette, France}
\affiliation{$^2$Aix Marseille Univ, CNRS, Centrale Marseille, Institut Fresnel, F-13013 Marseille, France}
\affiliation{$^3$American University of Technology, AUT Halat, Highway, Lebanon}
%%%%%%%%%%%%%%%%%%%%%%%%%%%%%%%%%%%%%%%%%%%%%%%%%%%%%%%%%%%%%%%%
\begin{abstract}
A new picture of the Josephson effect is devised. The radio-frequency (RF) signal, observed in a Josephson junction, is shown to stem from bound electrons, tunneling periodically through the insulating film. This holds also for the microwave mediated tunneling. The Josephson effect is found to be conditioned by the same prerequisite worked out previously for persistent currents, thermal equilibrium and occurence of superconductivity. The observed negative resistance behaviour is shown to originate from the interplay between normal and superconducting currents.
\end{abstract}
%%%%%%%%%%%%%%%%%%%%%%%%%%%%%%%%%%%%%%%%%%%%%%%%%%%%%%%%%%%%%%%%
%%%%%%%%%%%%%%%%%%%%%%%%%%%%%%%%%%%%%%%%%%%%%%%%%%%%%%%%%%%%%%%%
\maketitle
%%%%%%%%%%%%%%%%%%%%%%%%%%%%%%%%%%%%%%%%%%%%%%%%%%%%%%%%%%%%%%%%
	\section{Introduction}
\begin{figure}
\includegraphics*[height=4 cm,width=6 cm]{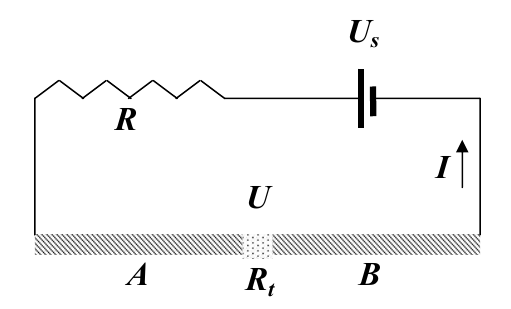}
\caption{Sketch of the electrical setup, operated to study the Josephson effect. The Josephson capacitor consists in two superconducting electrodes $A,B$ (hatched area) straddling an insulating film (dotted area); the insulator thickness has been hugely magnified for the reader's convenience. $U_s,U,R,R_t$ stand, respectively, for the constant applied bias, the voltage drop across the capacitor, a loading resistor inserted to measure the total current $I$ and the tunneling resistance, defined in section II.}
\label{jos1}
\end{figure}

%%%%%%%%%%%%%%%%%%%%%%%%%%%%%%%%%%%%%%%%%%%%%%%%%%%%%%%%%%%%%%%%
The Josephson effect was initially observed\cite{sha,sha2} in the kind of circuit sketched in Fig.\ref{jos1} and has kept arousing an unabated interest, in particular because of its relevance to electronic devices\cite{nag,gul} and quantum computation\cite{dou,dev,ydev}. For simplicity, both superconducting leads $A,B$ are assumed here to be made out of the same material. They are separated by a thin ($<10\texttt{\AA}$) insulating film, enabling electrons to tunnel through it. If $A,B$ were made of a normal metal, a constant current $I=\frac{U_s}{R+R_t}$ would flow through the circuit. Nevertheless, this simple setup has attracted considerable attention because of Josephson's predictions\cite{jos} :
%%%%%%%%%%%%%%%%%%%%%%%%%%%%%%%%%%%%%%%%%%%%%%%%%%%%%%%%%
	\begin{enumerate}
	\item 
	there should be $\left\langle I\right\rangle\neq 0$ for $\left\langle U\right\rangle=0$, which entails $\left|\frac{d\left\langle I\right\rangle}{d\left\langle U\right\rangle}(\left\langle U\right\rangle=0)\right|\rightarrow\infty$ ($\left\langle I\right\rangle,\left\langle U\right\rangle$ refer to time $t$ averaged values of $I(t),U(t)$);
	\item
	$I(t),U(t)$ should oscillate at frequency $\omega=\frac{2e\left\langle U\right\rangle}{\hbar}$ with $e$ being the electron charge.
\end{enumerate}\par
%%%%%%%%%%%%%%%%%%%%%%%%%%%%%%%%%%%%%%%%%%%%%%%%%%%%%%%%%
\begin{figure}
\includegraphics*[height=5 cm,width=5 cm]{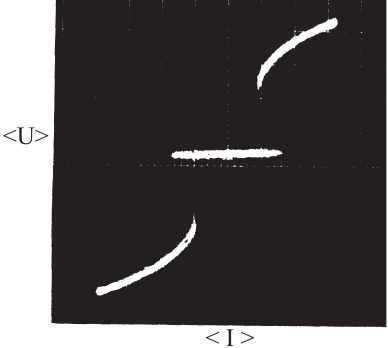}
\caption{Characteristic $I(U)$ recorded by Shapiro\cite{sha} and used here with APS permission; vertical scale is $58. 8 \mu V/cm$, horizontal scale is $130 nA/cm$.}
\label{sha1}
\end{figure}

%%%%%%%%%%%%%%%%%%%%%%%%%%%%%%%%%%%%%%%%%%%%%%%%%%%%%%%%%
However, claim $1$ seems to disagree with experimental data, reproduced in Fig.\ref{sha1}, because $\frac{d\left\langle I\right\rangle}{d\left\langle U\right\rangle}(\left\langle U\right\rangle=0)\approx .06\Omega^{-1}$ is seen  to be \textit{finite}.\par
%%%%%%%%%%%%%%%%%%%%%%%%%%%%%%%%%%%%%%%%%%%%%%%%%%%%%%%%%
	In addition, claim $1$ appears \textit{questionable} in view of the demurrals below :
%%%%%%%%%%%%%%%%%%%%%%%%%%%%%%%%%%%%%%%%%%%%%%%%%%%%%%%%%
\begin{itemize}
	\item 
	$\left\langle U\right\rangle=0$ implies that there is \textit{no electric field}, available to  accelerate the conduction electrons. Hence the \textit{finite} momentum, associated with the tunneling current $\left\langle I\right\rangle\neq 0$, has built up with \textit{no external force}, which \textit{violates} Newton's law;
	\item 
	since the electrons undergo \textit{no electric field}, it is hard to figure out why the tunneling current should flow into \textit{one} direction rather than the \textit{opposite} one;
	\item
	$\left\langle I\right\rangle\neq 0$ despite $U_s=\left\langle U\right\rangle=0$ entails that the $t$ averaged circulation of the electric field along the closed circuit, pictured in Fig.\ref{jos1}, equals $R\left\langle I\right\rangle\neq 0$ and thence the electric field is bound to be induced by a $t$ dependent magnetic field, according to the Faraday-Maxwell equation, in contradiction with the experimental setup in Fig.\ref{jos1}, involving \textit{no $t$ dependent} magnetic field.	
\end{itemize}
%%%%%%%%%%%%%%%%%%%%%%%%%%%%%%%%%%%%%%%%%%%%%%%%%%%%%%%%%
Besides a periodic signal was indeed observed\cite{sha,mcc}, but in the \textit{RF range}, i.e. $\omega<100MHz$, rather than in the \textit{microwave} one, i.e. $\omega>1GHz$, as inferred from Josephson's formula, given the measured $\left\langle U\right\rangle$ values.\par
%%%%%%%%%%%%%%%%%%%%%%%%%%%%%%%%%%%%%%%%%%%%%%%%%%%%%%%%%
	 Consequently, the numerous experimental data, documenting the electrodynamical behaviour of the Josephson junction\cite{nag,gul}, have been interpreted so far by resorting\cite{mcc} to a formula, relating $I(t),U(t)$ to Ginzburg and Landau's phase\cite{gin} $\Phi_{GL}$. However the time behaviour of $\Phi_{GL}(t)$ has been derived with help of a perturbation calculation\cite{jos,wer,lar,bar,lev}, which is well-suited to describe the \textit{random} tunneling of a single particle, either electron or Bogolyubov-Valatin excitation\cite{bar,lev}, but cannot account for the \textit{coherent} tunneling of \textit{bound} electrons, such as those making up the superconducting state\cite{sz4,sz5,sz7,sz8,sz9}, for some reason to be given below. Therefore, this work is rather intended at presenting an \textit{alternative} explanation of the Josephson effect, \textit{unrelated} to $\Phi_{GL}$,  by studying the \textit{time-periodic} tunneling motion\cite{schi} of bound electron pairs\cite{sz4,sz5,sz7,sz8,sz9} through the insulating barrier.\par
%%%%%%%%%%%%%%%%%%%%%%%%%%%%%%%%%%%%%%%%%%%%%%%%%%%%%%%%%
	The outline is as follows : the expression of the tunneling current, conveyed by independent electrons, is recalled in  section II, whereas the current carried by bound electrons is worked out in section III; this enables us to solve, in section IV, the electrodynamical equation of motion of the circuit, depicted in Fig.\ref{jos1}; sections V, VI deal respectively with the microwave mediated Josephson effect\cite{sha,sha2} and the negative resistance induced signal\cite{mcc}. The results are summarised in the conclusion.
%%%%%%%%%%%%%%%%%%%%%%%%%%%%%%%%%%%%%%%%%%%%%%%%%%%%%%%%%%%%%%%%
	\section{Random Tunneling}
%%%%%%%%%%%%%%%%%%%%%%%%%%%%%%%%%%%%%%%%%%%%%%%%%%%%%%%%%%%%%%%%
As in our previous work\cite{sz4,sz5,sz7,sz8,sz9,sz1,sz2,sz3}, the present analysis will proceed within the framework of the two-fluid model, for which the conduction electrons comprise superconducting and independent electrons, in respective concentration $c_s,c_n$. The superconducting and independent electrons are organized, respectively, as a many bound electron\cite{sz5} (MBE), BCS-like\cite{bcs} state, characterised by its chemical potential $\mu$, and a degenerate Fermi gas\cite{ash} of Fermi energy $E_F$. Assuming $U=U_A-U_B,eU>0$, the current, conveyed by the independent electrons, will flow from $A$ toward $B$ and there is $eU=E_F^A-E_F^B$, with $E_F^A,E_F^B$ being the Fermi energy in electrodes $A,B$, respectively. Hence, since the experiments are carried out at low temperature, the corresponding current density $j_n$ is inferred from the properties of the Fermi gas\cite{ash} to read 
%%%%%%%%%%%%%%%%%%%%%%%%%%%%%%%%%%%%%%%%%%%%%%%%%%%%%%%%%
%%%%%%%%%%%%%%%%%%%%%%%%%%%%%%%%%%%%%%%%%%%%%%%%%%%%%%%%%
%%%%%%%%%%%%%%%%%%%%%%%%%%%%%%%%%%%%%%%%%%%%%%%%%%%%%%
\begin{equation}
\label{jn}
j_n=\frac{e^2\rho(E_F^A)v_FT}{2}U\Rightarrow R_t\propto \frac{1}{\rho(E_F^A)}\quad,
\end{equation}
%%%%%%%%%%%%%%%%%%%%%%%%%%%%%%%%%%%%%%%%%%%%%%%%%%%%%%
with $\rho(E_F),v_F,T$ standing for the one-electron density of states at the Fermi level, the Fermi velocity and the one-electron transmission coefficient through the insulating barrier ($\Rightarrow 0<T<1$). Several remarks are in order, regarding Eq.(\ref{jn})
%%%%%%%%%%%%%%%%%%%%%%%%%%%%%%%%%%%%%%%%%%%%%%%%%%%%%%%%%%%%%%%%%%%%%
\begin{itemize}
	\item 
	Eq.(\ref{jn}) is seen to agree with the corresponding formula, available in textbooks\cite{bar,lev};
	\item 
	the independent electrons contribute thence the current $I_n(t)=U(t)/R_t$ to the total current $I(t)$. However, despite $I_n$ obeying Ohm's law, the tunneling electrons suffer no energy loss inside the insulating barrier; 
	\item
	because $c_n$ is expected to grow\cite{sz5} at the expense of $c_s$ with growing $\left| I\right|$, this implies that $\rho(E_F)$ and $R_t$ will, respectively, increase and decrease with increasing $\left| I\right|$. The negative resistance effect, addressed in section $6$, stems from this property.
\end{itemize}
%%%%%%%%%%%%%%%%%%%%%%%%%%%%%%%%%%%%%%%%%%%%%%%%%%%%%%%%%%%%%%%%
	\section{Coherent Tunneling}
%%%%%%%%%%%%%%%%%%%%%%%%%%%%%%%%%%%%%%%%%%%%%%%%%%%%%%%%%%%%%%%%
  Unlike the random diffusion of independent electrons across the insulating barrier, the tunneling motion of bound electrons takes place as a time-periodic oscillation to be analysed below. Their energy per unit volume $\mathcal{E}$ depends\cite{sz4} on $c_s$ only and is related to their chemical potential $\mu$ by $\mu=\frac{\partial\mathcal{E}}{\partial c_s}$. Before any electron crosses the barrier, the total energy of the whole bound electron system, including the leads $A,B$, reads
%%%%%%%%%%%%%%%%%%%%%%%%%%%%%%%%%%%%%%%%%%%%%%%%%%%%%%
\begin{equation}
\label{wh1}
\mathcal{E}_i=2\mathcal{E}(c_e)+ec_eU\quad,
\end{equation}
%%%%%%%%%%%%%%%%%%%%%%%%%%%%%%%%%%%%%%%%%%%%%%%%%%%%%%
with $c_e$ referring to the bound electron concentration at thermal equilibrium. Let $n>>1$ of bound electrons cross the barrier from $A$ toward $B$. The total energy becomes
%%%%%%%%%%%%%%%%%%%%%%%%%%%%%%%%%%%%%%%%%%%%%%%%%%%%%%
\begin{equation}
\label{wh2}
\mathcal{E}_f=\mathcal{E}(c_e+\frac{n}{V})+\mathcal{E}(c_e-\frac{n}{V})+e(c_e-\frac{n}{V})U\quad,
\end{equation}
%%%%%%%%%%%%%%%%%%%%%%%%%%%%%%%%%%%%%%%%%%%%%%%%%%%%%%
with $V$ being the volume, taken to be equal for both leads $A,B$. Energy conservation  requires $\mathcal{E}_i=\mathcal{E}_f$, which leads finally to
%%%%%%%%%%%%%%%%%%%%%%%%%%%%%%%%%%%%%%%%%%%%%%%%%%%%%%
\begin{equation}
\label{wh3}
n=\frac{eV}{\frac{\partial\mu}{\partial c_s}(c_e)}U\quad.
\end{equation}\par
%%%%%%%%%%%%%%%%%%%%%%%%%%%%%%%%%%%%%%%%%%%%%%%%%%%%%%
	The wave-functions $\varphi_i,\varphi_f$, associated with the twofold degenerate eigenvalue $\mathcal{E}_i=\mathcal{E}_f$, read  
%%%%%%%%%%%%%%%%%%%%%%%%%%%%%%%%%%%%%%%%%%%%%%%%%%%%%%
\begin{equation}
\label{eig1}
\begin{array}{l}
\varphi_i=\varphi_A(c_e)\otimes\varphi_B(c_e)\\
\varphi_f=\varphi_A(c_e-\frac{n}{V})\otimes\varphi_B(c_e+\frac{n}{V})
\end{array}\quad ,
\end{equation}
%%%%%%%%%%%%%%%%%%%%%%%%%%%%%%%%%%%%%%%%%%%%%%%%%%%%%%
with $\varphi(c_s)$ being the MBE, $c_s$ dependent eigenfunction\cite{bcs,sz5,sz9}. The coherent tunneling motion of $n$ electrons across the barrier is thence described by the wave-function $\psi(t)$, solution of the Schr\"{o}dinger equation 
%%%%%%%%%%%%%%%%%%%%%%%%%%%%%%%%%%%%%%%%%%%%%%%%%%%%%%
\begin{equation}
\label{sch1}
\begin{array}{c}
i\frac{\partial\psi}{\partial t}=H\psi\\
H=\omega_t\sigma_x\quad,\quad
\omega_t=\left\langle\varphi_i\left|V_b\right|\varphi_f\right\rangle
\end{array}\quad .
\end{equation}
%%%%%%%%%%%%%%%%%%%%%%%%%%%%%%%%%%%%%%%%%%%%%%%%%%%%%%
The Hamiltonian $H$ and the potential barrier $V_b$, hindering the electron motion through the Josephson junction and including the applied voltage $U$, are expressed in frequency unit, $\frac{V\mathcal{E}_i}{\hbar}$ is taken as the origin of energy, whereas $\psi$ and the Pauli matrix\cite{abr} $\sigma_x$ have been projected onto the basis $\{\varphi_i,\varphi_f\}$. The tunneling frequency $\omega_t$, taken to lie in the RF range, i.e. $\omega_t<100MHz$, as reported by Shapiro\cite{sha}, is realized to describe the tunneling motion of \textit{bound} electrons in a similar way as the matrix element $T_{kq}$ does for the \textit{random} tunneling of a \textit{single} electron in the mainstream view\cite{bar,lev}. Finally Eq.(\ref{sch1}) is solved\cite{abr} to yield
%%%%%%%%%%%%%%%%%%%%%%%%%%%%%%%%%%%%%%%%%%%%%%%%%%%%%%
  \begin{equation}
\label{psi1}
\psi(t)= \cos\left(\frac{\omega_t t}{2}\right)\varphi_i-i\sin\left(\frac{\omega_t t}{2}\right)\varphi_f\quad ,
\end{equation}
%%%%%%%%%%%%%%%%%%%%%%%%%%%%%%%%%%%%%%%%%%%%%%%%%%%%%%	
whence the charge $Q_s,-Q_s$, piling up in $A,B$ respectively, is inferred, thanks to Eq.\ref{wh3}, to read 
%%%%%%%%%%%%%%%%%%%%%%%%%%%%%%%%%%%%%%%%%%%%%%%%%%%%%%
$$Q_s(t)=-ne\left|\left\langle\psi(t)|\varphi_f\right\rangle\right|^2=C_eU\sin^2\left(\frac{\omega_t t}{2}\right)\quad,$$ 
%%%%%%%%%%%%%%%%%%%%%%%%%%%%%%%%%%%%%%%%%%%%%%%%%%%%%%
with the effective capacitance $C_e$ defined as
%%%%%%%%%%%%%%%%%%%%%%%%%%%%%%%%%%%%%%%%%%%%%%%%%%%%%%
		$$C_e=-\frac{e^2V}{\frac{\partial\mu}{\partial c_s}(c_e)}\quad .$$ \par
%%%%%%%%%%%%%%%%%%%%%%%%%%%%%%%%%%%%%%%%%%%%%%%%%%%%%%
Since $\frac{\partial\mu}{\partial c_s}<0$ has been shown to be a prerequisite for the  existence of persistent currents\cite{sz4}, thermal equilibrium\cite{sz5} and occurrence of superconductivity\cite{sz8,sz9}, it implies that $C_e>0$. In addition, given the estimate\cite{sz5} of $\frac{\partial\mu}{\partial c_s}$, it may take a very large value up to $C_e\approx 1F$. At last, by contrast with $I_n$ being incoherent, the bound electrons contribute an \textit{oscillating} current $I_s(t)={\dot Q}_s=\frac{dQ_s}{dt}$ to $I(t)$.\par
%%%%%%%%%%%%%%%%%%%%%%%%%%%%%%%%%%%%%%%%%%%%%%%%%%%%%%%%%%%%%%%%
	The marked difference between the \textit{random} diffusion current $I_n(t)$ and the \textit{time-periodic} one $I_s(t)$ ensues from the property that energy must be \textit{conserved} during tunneling. This is automatically ensured\cite{bar,lev} for an independent particle because its eigenenergy is defined uniquely all over the electrodes $A,B$ and the insulating barrier, whereas special care, as expressed in Eq.(\ref{wh3}), must be taken to enforce energy conservation for \textit{bound} electrons tunneling through a barrier. Unfortunately this \textit{crucial} constraint has been \textit{overlooked} in the mainstream analysis\cite{jos,wer,lar,bar,lev}.
%%%%%%%%%%%%%%%%%%%%%%%%%%%%%%%%%%%%%%%%%%%%%%%%%%%%%%%%%%%%%%%%
	\section{Electrodynamical Behaviour}
%%%%%%%%%%%%%%%%%%%%%%%%%%%%%%%%%%%%%%%%%%%%%%%%%%%%%%%%%%%%%%%%
	The total current $I(t)$ comprises $3$ contributions, namely $I_n=\frac{U}{R_t},I_s={\dot Q}_s$ and a component $C\dot U$, loading the Josephson capacitor ($C$ refers to its capacitance), so that the electrodynamical equation of motion reads
%%%%%%%%%%%%%%%%%%%%%%%%%%%%%%%%%%%%%%%%%%%%%%%%%%%%%%
$$U_s=U+RI\quad,\quad I=\frac{U}{R_t}+{\dot Q}_s+C\dot U\quad,$$
%%%%%%%%%%%%%%%%%%%%%%%%%%%%%%%%%%%%%%%%%%%%%%%%%%%%%%
which is finally recast into 
%%%%%%%%%%%%%%%%%%%%%%%%%%%%%%%%%%%%%%%%%%%%%%%%%%%%%%
\begin{equation}
\label{eqm}
\dot U=\frac{U_s-U\left(1+\frac{R}{R_t}+\frac{RC_e\omega_t}{2}\sin\left(\omega_t t\right)\right)}{R\left(C+C_e\sin^2\left(\frac{\omega_t t}{2}\right)\right)}\quad.
\end{equation}
%%%%%%%%%%%%%%%%%%%%%%%%%%%%%%%%%%%%%%%%%%%%%%%%%%%%%%
It is worth noticing that, due to $\left|\frac{C_e}{C}\right|>>1$, the denominator in the right-hand side of Eq.(\ref{eqm}) would vanish for $C_e<0$, at some $t$ value, so that Eq.(\ref{eqm}) cannot be solved unless $C_e>0\Rightarrow \frac{\partial\mu}{\partial c_s}<0$, which confirms a \textit{previous}\cite{sz4,sz5,sz7,sz8,sz9} conclusion, derived independently.\par
%%%%%%%%%%%%%%%%%%%%%%%%%%%%%%%%%%%%%%%%%%%%%%%%%%%%%%
%%%%%%%%%%%%%%%%%%%%%%%%%%%%%%%%%%%%%%%%%%%%%%%%%%%%%%%%%
\begin{figure}
\includegraphics*[height=12 cm,width=8 cm]{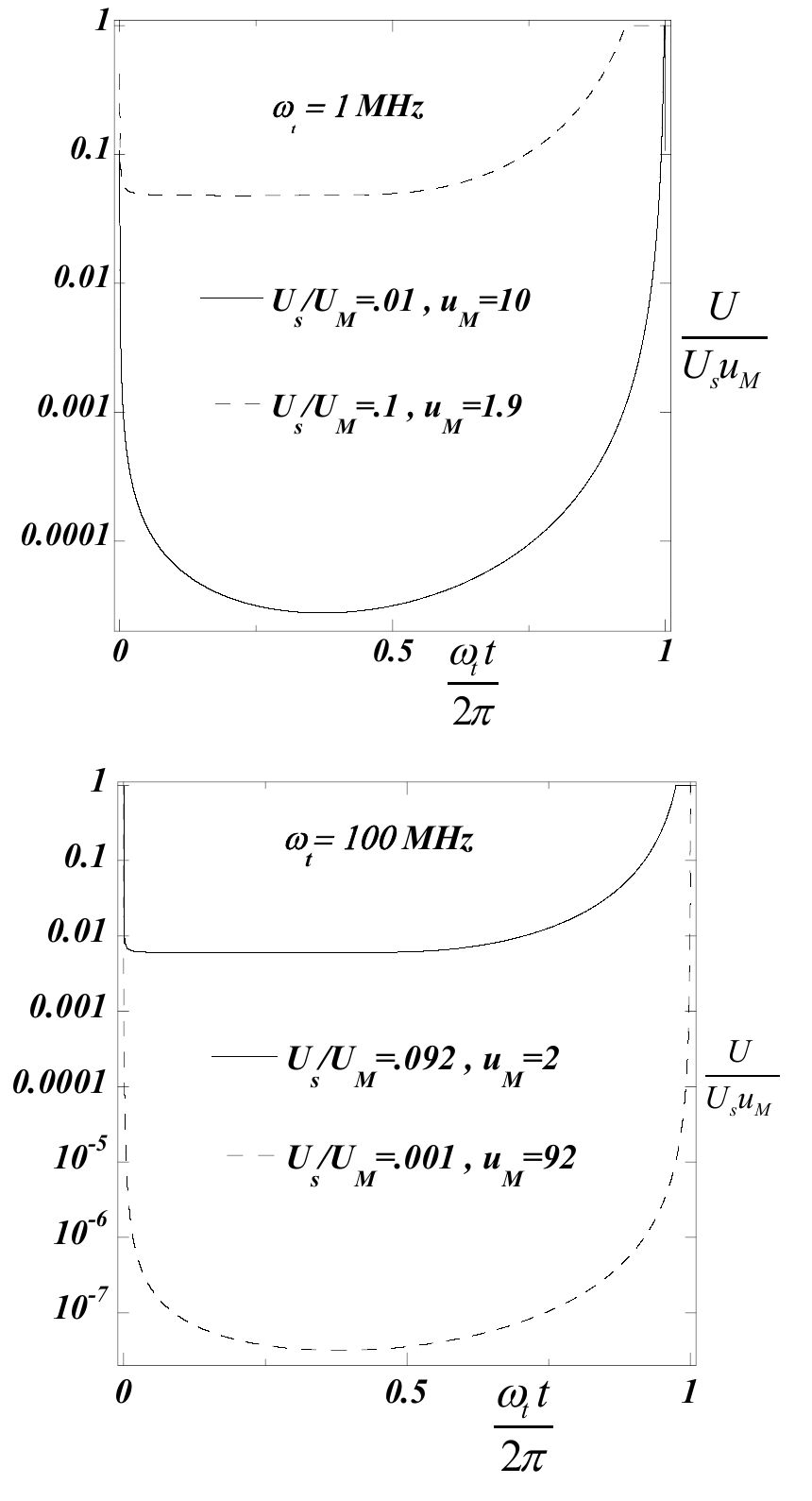}
\caption{Semi-logarithmic plots of the periodic solution $U(t)$ of Eq.(\ref{eqm}), calculated for $\omega_t=1MHz,100MHz$ and $I_M=1mA,0.1mA$; $U_M=\left(R+R_n\right)I_M$ and $u_M$ is the maximum value of $\frac{\left|U\left(t\in\left[0,\frac{2\pi}{\omega_t}\right]\right)\right|}{U_s}$.}
\label{per}
\end{figure}
%%%%%%%%%%%%%%%%%%%%%%%%%%%%%%%%%%%%%%%%%%%%%%%%%%%%%%%%%
	$\left|\frac{\partial\mu}{\partial c_s}\right|$ is expected\cite{sz5} to increase with increasing $\left|I\right|$ and is no longer defined for $\left|I\right|>I_M$, the maximum value of the bound electron current, because the sample goes thereby normal. Consequently for practical purposes, Eq.(\ref{eqm}) has been solved by assuming $R_t\left(\left|I\right|\leq I_M\right)=R_0g\left(\left|\frac{I}{I_M}\right|\right)+R_n$, $R_t\left(\left|I\right|>I_M\right)=R_n$ with $\frac{R_0}{R_n}>>1$, $C_e\left(\left|I\right|\leq I_M\right)=C_0g\left(\left|\frac{I}{I_M}\right|\right)$, $C_e\left(\left|I\right|>I_M\right)=0$ with $\frac{C_0}{C}>>1$, and $g(x)=1-x^2$. Regardless of the initial condition $U(0)$, the solution $U(t)$ of Eq.(\ref{eqm}) becomes time-periodic, i.e. $U\left(t\right)=U\left(t+\frac{2\pi}{\omega_t}\right),\forall t$, after a short transient regime.\par
%%%%%%%%%%%%%%%%%%%%%%%%%%%%%%%%%%%%%%%%%%%%%%%%%%%%%%
	Eq.(\ref{eqm}) has been solved with the assignments $C=1pF,C_0=1mF,R=10\Omega,R_n=100\Omega,R_0=10K\Omega$, and the corresponding $U(t)$ have been plotted in Fig.\ref{per}. The large slope $\left|\frac{dU}{dt}(0)\right|>>1$ stems from $\frac{C_0}{C}>>1$. Since no experimental data of $U(t)$ have been reported in the literature to the best of our knowledge, no comparison between observed and calculated results can be done. Nevertheless, the large $u_M>>1$ values, seen in Fig.\ref{per}, have been indeed observed\cite{sha}. Likewise, the calculated $u_M$ values have been found to increase very steeply with $U_s$ decreasing toward $0$. Hence the thermal noise, generated by the $U_s$ source, will suffice even at $U_s=0$ to give rise to sizeable $u_M$, which is likely to be responsible for the \textit{misconception}\cite{jos}, conveyed by hereabove mentioned claim $1$ . As a matter of fact, the noisy behaviour of the circuit sketched in Fig.\ref{jos1} has been reported\cite{sha}.\par
\begin{figure}
\includegraphics*[height=12 cm,width=8 cm]{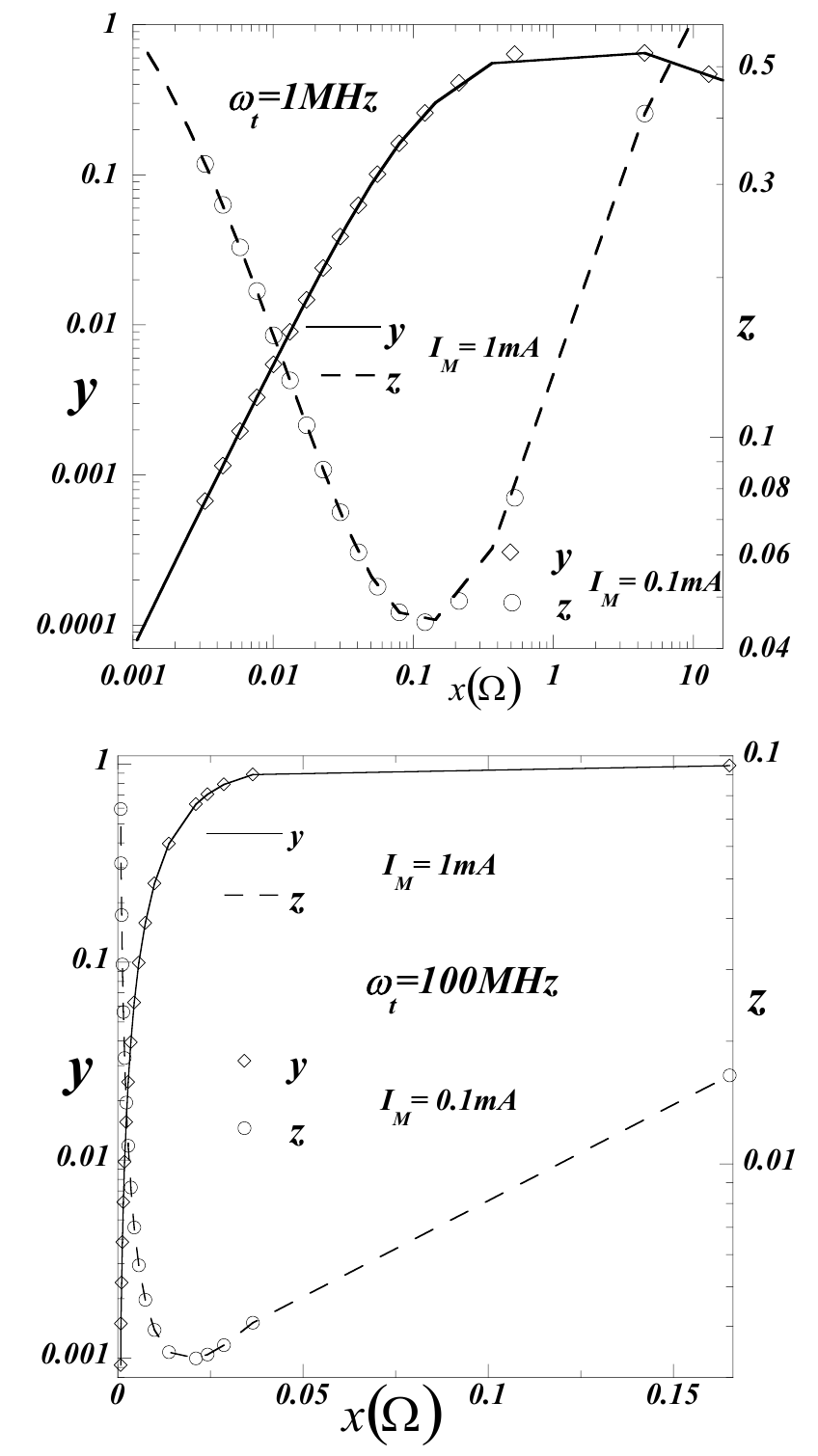}
\caption{Logarithmic and semi-logarithmic plots of the characteristics $I(U)$, calculated for $\omega_t=1MHz,100MHz$, respectively, and $I_M=1mA,0.1mA$, with $x=\frac{\left\langle U\right\rangle}{I_M}$, $y=\frac{\left\langle I\right\rangle}{I_M}$, $z=\frac{\left\langle U\right\rangle}{U_s}$.}
\label{char}
\end{figure}

%%%%%%%%%%%%%%%%%%%%%%%%%%%%%%%%%%%%%%%%%%%%%%%%%%%%%% 
	The characteristics $I(U)$, plotted in Fig.\ref{char}, have been reckoned as
%%%%%%%%%%%%%%%%%%%%%%%%%%%%%%%%%%%%%%%%%%%%%%%%%%%%%%%%%%%%%%%%
	$$\left\langle f\right\rangle=\frac{\omega}{2\pi}\int_0^{\frac{2\pi}{\omega}}f(u)du\quad,$$
%%%%%%%%%%%%%%%%%%%%%%%%%%%%%%%%%%%%%%%%%%%%%%%%%%%%%%%%%%%%%%%%
with $f=U,I$. In all cases, there is $\left\langle I\right\rangle(0)=0$ with \textit{finite} $\frac{d\left\langle I\right\rangle}{d\left\langle U\right\rangle}(0)$ in agreement with the experimental data in Fig.\ref{sha1}. However the slope $\frac{d\left\langle I\right\rangle}{d\left\langle U\right\rangle}(0)$, calculated for $\omega_t=100MHz$, is much larger than the one at $\omega_t=1MHz$.\par
%%%%%%%%%%%%%%%%%%%%%%%%%%%%%%%%%%%%%%%%%%%%%%%%%%%%%%%%%%%%%%%%
	Noteworthy is that there are no observed $\left\langle I\right\rangle$ data in Fig.\ref{sha1} over a broad $\left\langle U\right\rangle$ range, starting from $\left\langle U\right\rangle\approx 0$ up to a value big enough for the sample to go into the normal state, characterised by constant $I=I_n>I_M$. This feature might result\cite{sha} from $U_s\propto\sin(\omega_p t)$ with $\omega_p=60Hz$. Thus since the tunneling frequency $\omega_t$ is expected to decrease exponentially\cite{bar,lev,schi} with increasing $n$ and thence $U$, this entails that the signal could indeed no longer be observed for $\omega_t<\omega_p$. Likewise, the observed frequency modulation\cite{sha2,bar,lev}, i.e. $\omega_t(n)$ is time-periodic, ensues from $n(t)\propto U(t)$ being time-periodic too (see Eq.(\ref{wh3})). At last, it is in order to realize that the characteristics $I(U)$ is anyhow \textit{not} an intrinsic property of the Josephson junction, because it depends on $R$, as seen in Eq.(\ref{eqm}).
%%%%%%%%%%%%%%%%%%%%%%%%%%%%%%%%%%%%%%%%%%%%%%%%%%%%%%%%%
%%%%%%%%%%%%%%%%%%%%%%%%%%%%%%%%%%%%%%%%%%%%%%%%%%%%%%%%%
%%%%%%%%%%%%%%%%%%%%%%%%%%%%%%%%%%%%%%%%%%%%%%%%%%%%%%%%%%%%%%%%
	\section{Microwave Mediated Tunneling}
\begin{figure}
\includegraphics*[height=7 cm,width=7 cm]{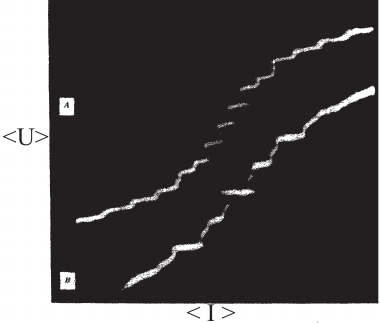}
\caption{Characteristics $I(U)$, recorded by Shapiro\cite{sha} (used here with APS permission) at $9.3$GHz for A (vertical scale is 58.8 pV/cm, horizontal scale is 67 nA/cm) and $24.85$GHz for B (vertical scale is 50 pV/cm, horizontal scale is 50 pA/cm).}
\label{step}
\end{figure}

%%%%%%%%%%%%%%%%%%%%%%%%%%%%%%%%%%%%%%%%%%%%%%%%%%%%%%%%%%%%%%%%	
	By irradiating the Josephson junction, depicted in Fig.\ref{jos1}, with an electromagnetic microwave of frequency $\omega$, Shapiro observed\cite{sha} the step-like characteristic $I(U)$, recalled in Fig.\ref{step}. The discontinuities of $\frac{d\left\langle I\right\rangle}{d\left\langle U\right\rangle}$, showing up at $\left\langle U\right\rangle=\frac{m\hbar\omega}{2e}$ with $m>0$ being an integer, brought forward a cogent proof that the MBE state comprises an \textit{even} number of electrons. In order to explain this experimental result, let us begin with studying the microwave induced tunneling of \textit{one} bound electron pair across the $U_m=\frac{m\hbar\omega}{2e}$ biased barrier. The corresponding Hilbert space, describing the system \textit{before} and \textit{after} crossing, is subtended by the basis $\left\{\varphi_i=\varphi_A(c_e)\otimes\varphi_B(c_e),\varphi_1=\varphi_A(c_e+\frac{2}{V})\otimes\varphi_B(c_e-\frac{2}{V})\right\}$ of respective energies $V\mathcal{E}_i,V\mathcal{E}_i+m\hbar\omega$. The tunneling motion of one electron pair is then described by $\psi_0(t)$, solution of the Schr\"{o}dinger equation 
%%%%%%%%%%%%%%%%%%%%%%%%%%%%%%%%%%%%%%%%%%%%%%%%%%%%%%
\begin{equation}
\label{sch2}
\begin{array}{c}
i\frac{\partial\psi_0}{\partial t}=H_0(t)\psi\\
H_0=m\omega\sigma_z+2\left(\omega_t+\omega_r\sin\left(\omega t\right)\right)\sigma_x
\end{array}\quad .
\end{equation}
%%%%%%%%%%%%%%%%%%%%%%%%%%%%%%%%%%%%%%%%%%%%%%%%%%%%%%
The Hamiltonian $H_0$ is expressed in frequency unit, $\frac{V\mathcal{E}_i}{\hbar}+\frac{m\omega}{2}$ is taken as the origin of energy, $\omega_r$ stands for the dipolar, off-diagonal matrix element\cite{boy} (the microwave power is $\propto\omega^2_r$), and $\sigma_z,\sigma_x$ are Pauli's matrices\cite{abr}, projected onto $\left\{\varphi_i,\varphi_1\right\}$. It is worth pointing out that Eq.(\ref{sch2}) could be readily solved like Eq.(\ref{sch1}), if $H_0$ were $t$ independent. Accordingly, in order to get rid of the $t$ dependence of $H_0$, we shall take advantage of a procedure devised for nonlinear optics\cite{ja4,ja5}.\par
%%%%%%%%%%%%%%%%%%%%%%%%%%%%%%%%%%%%%%%%%%%%%%%%%%%%%%
	To that end, $H_0$ is first recast into 
%%%%%%%%%%%%%%%%%%%%%%%%%%%%%%%%%%%%%%%%%%%%%%%%%%%%%%
\begin{equation}
\label{pq}
H_0=P_0+f(t)\sigma_x\quad,
\end{equation}
%%%%%%%%%%%%%%%%%%%%%%%%%%%%%%%%%%%%%%%%%%%%%%%%%%%%%%
for which $P_0=m\omega\sigma_z+2\omega_t\sigma_x$ is a Hermitian, $2\times 2$, $t$ independent matrix, such that $\left(P_0\right)_{1,1}+\left(P_0\right)_{2,2}=0$, $\left(P_0\right)_{2,2}-\left(P_0\right)_{1,1}=m\omega$, and $f(t)=\omega_r\sin\left(\omega t\right)$ is a real function of period $=\frac{2\pi}{\omega}$, having the dimension of a frequency, such that $\left\langle f\right\rangle=\int_0^{\frac{2\pi}{\omega}}f(t)dt=0$. Then $H_0$ is projected onto $\left\{\psi_-,\psi_+\right\}$, the eigenbasis of $P_0$
%%%%%%%%%%%%%%%%%%%%%%%%%%%%%%%%%%%%%%%%%%%%%%%%%%%%%%
\begin{equation}
\label{tht}
G=TH_0T^{-1}=\epsilon\sigma_z+d(t)\sigma_z+g(t)\sigma_x\quad.
\end{equation}
%%%%%%%%%%%%%%%%%%%%%%%%%%%%%%%%%%%%%%%%%%%%%%%%%%%%%%
$T$ is the unitary transfer matrix from $\left\{\varphi_i,\varphi_1\right\}$ to $\left\{\psi_-,\psi_+\right\}$ and $\sigma_z,\sigma_x$ have been projected onto $\left\{\psi_-,\psi_+\right\}$. The corresponding eigenvalues are $\mp\frac{\epsilon}{2}$ with $\epsilon=\sqrt{(m\omega)^2+\omega_t^2}\approx m\omega$ because of $\omega_t<<\omega$, while the real functions $d(t),g(t)$ have the same properties as $f(t)$ in Eq.(\ref{pq}). Let us now introduce\cite{ja4,ja5} the unitary transformation $R_1(t)$, operating in the Hilbert space, subtended by $\left\{\psi_-,\psi_+\right\}$
%%%%%%%%%%%%%%%%%%%%%%%%%%%%%%%%%%%%%%%%%%%%%%%%%%%%%%
\begin{equation}
\label{uni}
R_1(t)=e^{i\Phi(t)}\left|\psi_-\right\rangle\left\langle\psi_-\right|+e^{-i\Phi(t)}\left|\psi_+\right\rangle\left\langle\psi_+\right|\quad ,
\end{equation}
%%%%%%%%%%%%%%%%%%%%%%%%%%%%%%%%%%%%%%%%%%%%%%%%%%%%%%
with the dimensionless $\Phi(t)=\frac{\omega t}{2}-\int_0^t d(u)du$. We then look for $\psi_1=R_1^{-1}\psi_0$, solution of the Schr\"{o}dinger equation 
%%%%%%%%%%%%%%%%%%%%%%%%%%%%%%%%%%%%%%%%%%%%%%%%%%%%%%
\begin{equation}
\label{sch3}
\begin{array}{c}
i\frac{\partial\psi_1}{\partial t}=H_1\psi_1\quad,H_1=R_1^{-1}GR_1-iR_1^{-1}\dot R_1\quad\\
H_1=P_1+\Re(z_1(t))\sigma_x+\Im(z_1(t))\sigma_y\\
P_1=\epsilon\sigma_z+2\omega_1\sigma_x
\end{array}\quad,
\end{equation}
%%%%%%%%%%%%%%%%%%%%%%%%%%%%%%%%%%%%%%%%%%%%%%%%%%%%%%
for which the Hermitian $2\times 2$ matrix $P_1$ has the same properties as $P_0$ in Eq.(\ref{pq}), except for $\left(P_1\right)_{2,2}-\left(P_1\right)_{1,1}\approx(m-1)\omega,\left(P_1\right)_{2,1}=\omega_1=\omega_r/2$ instead of $\left(P_0\right)_{2,2}-\left(P_1\right)_{1,1}=m\omega,\left(P_0\right)_{2,1}=\omega_t$, the Pauli matrices $\sigma_z,\sigma_x,\sigma_y$ have been projected onto $\left\{\psi_-,\psi_+\right\}$, and $\Re(z_1(t)),\Im(z_1(t))$ which are the real and imaginary parts of the complex function $z_1(t)$, have the same properties as $f(t)$ in Eq.(\ref{pq}). Consequently, iterating this procedure $m$ of times yields finally
%%%%%%%%%%%%%%%%%%%%%%%%%%%%%%%%%%%%%%%%%%%%%%%%%%%%%%
\begin{equation}
\label{sch4}
\begin{array}{c}
i\frac{\partial\psi_m}{\partial t}=H_m\psi_m\\
H_m=P_m+\Re(z_m(t))\sigma_x+\Im(z_m(t))\sigma_y\\
P_m=\eta\sigma_z+2\left(\Re(\omega_m)\sigma_x+\Im(\omega_m)\sigma_y\right)
\end{array}\quad,
\end{equation}
%%%%%%%%%%%%%%%%%%%%%%%%%%%%%%%%%%%%%%%%%%%%%%%%%%%%%%
for which the Pauli matrices $\sigma_z,\sigma_x,\sigma_y$ have been projected onto the eigenbasis of $P_m$, $\left\{\psi_-,\psi_+\right\}$, and $\eta\approx0$, $|\omega_m|<<\omega_r$. The Fourier series $\Re(z_m(t)),\Im(z_m(t))$ of fundamental frequency $\omega$ play no role, because the resonance condition\cite{abr} $\left|\left(P_m\right)_{1,1}-\left(P_m\right)_{2,2}\right|=\omega$ is not fulfilled due to $\left|\left(P_m\right)_{1,1}-\left(P_m\right)_{2,2}\right|=|\eta|<<\omega$, so that Eq.(\ref{sch4}) is finally solved, similarly to Eq.(\ref{sch1}), to give
%%%%%%%%%%%%%%%%%%%%%%%%%%%%%%%%%%%%%%%%%%%%%%%%%%%%%%
$$\psi_m= \cos\left(\frac{|\omega_m| t}{2}\right)\psi_--i\sin\left(\frac{|\omega_m| t}{2}\right)\psi_+\quad .$$
%%%%%%%%%%%%%%%%%%%%%%%%%%%%%%%%%%%%%%%%%%%%%%%%%%%%%% 
The solution of Eq.(\ref{sch2}) is thereby inferred to read
%%%%%%%%%%%%%%%%%%%%%%%%%%%%%%%%%%%%%%%%%%%%%%%%%%%%%%
$$\psi_0(t)= \left(\prod_{i=1,m}R_i(t)\right)\psi_m(t)\quad .$$
%%%%%%%%%%%%%%%%%%%%%%%%%%%%%%%%%%%%%%%%%%%%%%%%%%%%%% 
$U_m$ can be fitted to get $\eta=0$. Thus, for the sake of illustration, calculated $|\omega_m|$ and $\delta_m=1-\frac{2eU_m}{m\hbar\omega}$ are indicated in table \ref{tab}. As expected, $|\omega_m|$ decreases steeply with increasing $m$ but, remarkably enough, $|\omega_{2m+1}|$ decreases more slowly than $|\omega_{2m}|$, all the more so since $\omega_t$ is weaker. This property ensues\cite{abr,boy} from $\omega_{2m}=0,\forall m$ for $\omega_t=0$.\par
%%%%%%%%%%%%%%%%%%%%%%%%%%%%%%%%%%%%%%%%%%%%%%%%%%%%%%%%%
\begin{table}
\caption{calculated $|\omega_m|,\delta_m$ values  with $\omega=10GHz$, $\omega_r=100MHz$ and $\omega_t=100MHz,1MHz$.}
\label{tab}
%%%%%%%%%%%%%%%%%%%%%%%%%%%%%%%%%%%%%%%%%%%%%%%%%%%%%%%%%
\begin{center}
%%%%%%%%%%%%%%%%%%%%%%%%%%%%%%%%%%%%%%%%%%%%%%%%%%%%%%%%%
\begin{tabular}{|c|cc|cc|}
\hline
& $\omega_t=$ & $100MHz$ & $\omega_t=$ & $1MHz$\\
\hline
$m$ & $\frac{|\omega_m|}{\omega_r}$ & $\delta_m$ & $\frac{|\omega_m|}{\omega_r}$ & $\delta_m$\\
\hline 
$1$ & $0.5$ & $2\times 10^{-4}$ & $0.5$ & $2\times 10^{-8}$ \\
\hline
$2$ & $5\times 10^{-5}$ & $8\times 10^{-5}$ & $5\times 10^{-7}$ & $3\times 10^{-5}$ \\
\hline
$3$ & $3\times 10^{-6}$ & $3\times 10^{-5}$ & $3\times 10^{-6}$ & $8\times 10^{-6}$ \\
\hline
$4$ & $10^{-10}$ & $2\times 10^{-5}$ & $6\times 10^{-13}$ & $4\times 10^{-6}$ \\
\hline
$5$ & $5\times 10^{-12}$ & $10^{-5}$ & $5\times 10^{-12}$ & $3\times 10^{-6}$ \\
\hline
$6$ & $ 5\times 10^{-13}$ & $7\times 10^{-6}$ & $5\times 10^{-13}$ & $2\times 10^{-6}$ \\
\hline
$7$ & $5\times 10^{-13}$ & $5\times 10^{-6}$ & $5\times 10^{-13}$ & $10^{-6}$ \\
\hline
$8$ & $7\times 10^{-13}$ & $4\times 10^{-6}$ & $7\times 10^{-13}$ & $9\times 10^{-7}$ \\
\hline 
$9$ & $10^{-12}$ & $3\times 10^{-6}$ & $10^{-12}$ & $7\times 10^{-7}$ \\
\hline
$10$ & $3\times 10^{-12}$ & $3\times 10^{-6}$ & $3\times 10^{-12}$ & $6\times 10^{-7}$ \\
\hline
$11$ & $4\times 10^{-12}$ & $2\times 10^{-6}$ & $4\times 10^{-12}$  &  $5\times 10^{-7}$ \\
\hline  
\end{tabular}
%%%%%%%%%%%%%%%%%%%%%%%%%%%%%%%%%%%%%%%%%%%%%%%%%%%%%%%%%
\end{center}
\end{table}
%%%%%%%%%%%%%%%%%%%%%%%%%%%%%%%%%%%%%%%%%%%%%%%%%%%%%%%%%
%%%%%%%%%%%%%%%%%%%%%%%%%%%%%%%%%%%%%%%%%%%%%%%%%%%%%%%%%
%%%%%%%%%%%%%%%%%%%%%%%%%%%%%%%%%%%%%%%%%%%%%%%%%%%%%%%%% 
	Let us neglect $\frac{2eU_m}{V\mathcal{E}_i}<10^{-20}$, so that the energy of $\psi_0$ is taken to be constant and equal to $V\mathcal{E}_i$. The coherent tunneling of $n>>2$ of bound electrons will thence be described by Eq.(\ref{psi1}), except for $\left\{\psi_0,\varphi_f\right\}$, $\left\langle U\right\rangle-U_m$, $\left\langle I_m\right\rangle$, showing up instead of $\left\{\varphi_i,\varphi_f\right\}$, $\left\langle U\right\rangle$, $\left\langle I\right\rangle$, respectively, which entails that $\left\langle I_m\right\rangle\left(\left\langle U\right\rangle-U_m\right)=\left\langle I\right\rangle\left(\left\langle U\right\rangle\right)$, as illustrated by Fig.\ref{char}. Likewise, the contributions $\left\langle I_{m=1,2,3...}\right\rangle$ will add up together to give the step-like characteristic $I(U)$, recalled in Fig.\ref{step}. At last, Shapiro noticed\cite{sha} that some contributions $\left\langle I_{m}\right\rangle$ were missing in Fig.\ref{step}. As explained above in section $4$, this might result from the corresponding $|\omega_m|<\omega_p$ and thence would confirm $\omega_t<<\omega$.
%%%%%%%%%%%%%%%%%%%%%%%%%%%%%%%%%%%%%%%%%%%%%%%%%%%%%%%%%
\begin{figure}
\includegraphics*[height=5 cm,width=5 cm]{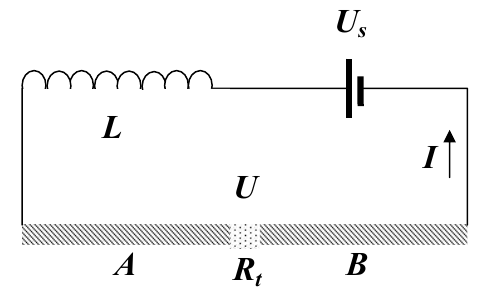}
\caption{Sketch of the electrical setup, displaying the negative resistance behaviour. $L$ refers to the self-inductance of the coil.}
\label{jos2}
\end{figure}
%%%%%%%%%%%%%%%%%%%%%%%%%%%%%%%%%%%%%%%%%%%%%%%%%%%%%%%%%
%%%%%%%%%%%%%%%%%%%%%%%%%%%%%%%%%%%%%%%%%%%%%%%%%%%%%%%%%%%%%%%%%
				\section{Negative Resistance}
\begin{figure}
\includegraphics*[height=5.5 cm,width=6 cm]{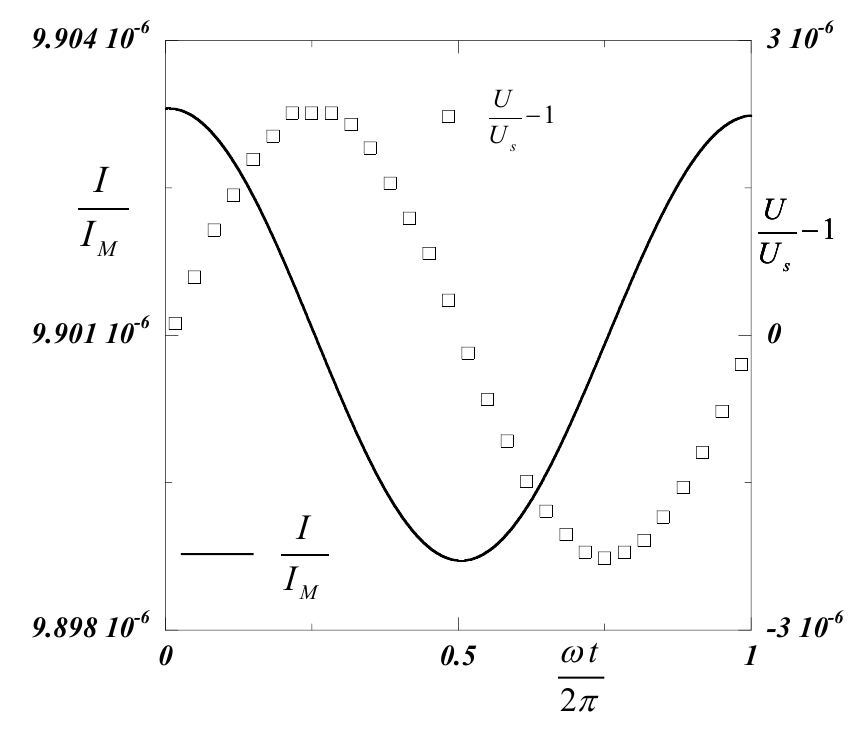}
\caption{Plots of the periodic solution $I(t),U(t)$ of Eq.(\ref{neg}), reckoned with $U_s=10\mu V,I_M=0.1mA,L=1\mu H,C=100pF,\omega=100MHz$.}
\label{sin}
\end{figure}

%%%%%%%%%%%%%%%%%%%%%%%%%%%%%%%%%%%%%%%%%%%%%%%%%%%%%%%%%%%%%%%%%
Signals $U(t),I(t)\propto\sin(\omega t)$, with the RF frequency $\omega$ defined by the resonance condition $LC\omega^2=1$, have been observed\cite{mcc} in the kind of setup, sketched in Fig.\ref{jos2}. Due to $\omega\neq\omega_t$, the bound electron tunneling  plays no role and the oscillation rather stems from $R_t(I)$ decreasing\cite{sz5} down to $R_n$ with $|I|$ increasing up to $I_M$, as indicated in section 4. Accordingly, since the voltage drop across the coil is equal to $L\dot I$, the electrodynamical equation of motion reads
%%%%%%%%%%%%%%%%%%%%%%%%%%%%%%%%%%%%%%%%%%%%%%%%%%%%%%
\begin{equation}
\label{neg}
 I=\frac{U}{R_t}+C\dot U\Rightarrow\ddot U=\omega^2(U_s-U)-\frac{\dot U}{R_tC}\quad .
\end{equation}
%%%%%%%%%%%%%%%%%%%%%%%%%%%%%%%%%%%%%%%%%%%%%%%%%%%%%%
Linearising Eq.(\ref{neg}) around the fixed point $U_0=U_s\Rightarrow I_0=\frac{U_s}{R_t(I_0)}$ yields the differential equation
%%%%%%%%%%%%%%%%%%%%%%%%%%%%%%%%%%%%%%%%%%%%%%%%%%%%%%
\begin{equation}
\label{lin}
 \ddot U=-\omega^2U-\frac{\dot U}{R_eC}\quad ,
\end{equation}
%%%%%%%%%%%%%%%%%%%%%%%%%%%%%%%%%%%%%%%%%%%%%%%%%%%%%%
with the effective resistance $R_e$, defined by $R_e=R_t(I_0)+I_0\frac{dR_t}{dI}(I_0)$. Due to $\frac{dR_t}{dI}<0$, the fixed point may be unstable in case of negative resistance $R_e<0$, which will give rise to an oscillating solution of Eq.(\ref{neg}), $U(t)\propto\sin(\omega t)$. As a matter of fact, integrating Eq.(\ref{neg}) leads to the sine-wave, depicted in Fig.\ref{sin}. Note that, unlike $U(t)$ in Fig.\ref{per}, every harmonic  $\propto\sin(m\omega t)$ with $m>1$ is efficiently smothered by the resonating $L,C$ circuit due to $LC(m\omega)^2\ne 1$ for $m>1$. At last, we have checked that Eq.(\ref{neg}) has no sine-wave solution for $\frac{R_0}{R_n}<50$ or $U_s>R_nI_M$, because those inequalities entail that $R_e>0$, which corresponds to a stable fixed point of Eq.(\ref{lin}).
%%%%%%%%%%%%%%%%%%%%%%%%%%%%%%%%%%%%%%%%%%%%%%%%%%%%%%%%%
%%%%%%%%%%%%%%%%%%%%%%%%%%%%%%%%%%%%%%%%%%%%%%%%%%%%%%%%% 
				\section{Conclusion}
%%%%%%%%%%%%%%%%%%%%%%%%%%%%%%%%%%%%%%%%%%%%%%%%%%%%%%%%%%%%%%%%%
All experimental results\cite{sha}, illustrating the Josephson effect, have been accounted for on the basis of bound electrons tunneling periodically across the insulating barrier. Likewise, the very existence of the Josephson effect has been shown to be conditioned by $\frac{\partial\mu}{\partial c_s}<0$, which had previously been recognized as a prerequisite for persistent currents\cite{sz4}, thermal equilibrium\cite{sz5}, a stable superconducting phase\cite{sz8} and a second order transition\cite{sz9}, occuring at the critical temperature $T_c$  too. The negative resistance feature\cite{mcc} has been ascribed to the tunneling resistance of independent electrons decreasing with increasing current, flowing through the superconducting electrodes, which confirms the validity of an analysis of the superconducting-normal transition\cite{sz5}.\par
%%%%%%%%%%%%%%%%%%%%%%%%%%%%%%%%%%%%%%%%%%%%%%%%%%%%%%%%%
	By contrast with this work, $I_n(t),I_s(t)$ are dealt with on the \textit{same footing} in the mainstream view\cite{jos,wer,lar,bar,lev}, both resulting from the tunneling of independent particles, obeying \textit{Fermi-Dirac} statistics. The only difference appears to be the one-particle density of states, namely either that associated with normal electrons for $I_n$ or Bogoliubov-Valatin\cite{par,sch} excitations for $I_s$.\par
%%%%%%%%%%%%%%%%%%%%%%%%%%%%%%%%%%%%%%%%%%%%%%%%%%%%%%%%%
	The coherent tunneling of bound electrons is thus concluded to be the very signature of the Josephson effect. Furthermore it has two noticeable properties :
%%%%%%%%%%%%%%%%%%%%%%%%%%%%%%%%%%%%%%%%%%%%%%%%%%%%%%%%%
\begin{itemize}
	\item
	since coherent tunneling has been ascribed in the third section to the properties of a MBE state, the time-periodic tunneling of bound electrons through a thin insulating barrier might be observed on a Josephson capacitor, for which the superconducting electrodes $A,B$ would be replaced by magnetic (ferromagnetic or antiferromagnetic) metals\cite{lev};
	\item
		the coherent tunneling motion seems to have no counterpart in the microscopic realm. For instance, the electrons, involved in a covalent bond, cannot tunnel between the two bound atoms because of their thermal relaxation toward the bonding groundstate. As for the Josephson effect, the bonding eigenfunction and its associated energy would read $\varphi_b=\frac{\varphi_i+\varphi_f}{\sqrt{2}}$ and $V\mathcal{E}_i-\frac{\hbar\omega_t}{2}$, respectively, but the relaxation from the tunneling state $\psi(t)$ in Eq.(\ref{sch1}) toward $\varphi_b$ might occur only inside the insulating barrier, which is impossible because the valence band, being fully occupied, can thence accomodate no additional electron.
\end{itemize}
%%%%%%%%%%%%%%%%%%%%%%%%%%%%%%%%%%%%%%%%%%%%%%%%%%%%%%%%%%%%%%%%%%%%%%%%%

%%%%%%%%%%%%%%%%%%%%%%%%%%%%%%%%%%%%%%%%%%%%%%%%%%%%%%%%%%%%%%%%
\end{document}